\documentclass[preprint,showpacs,showkeywords,preprintnumbers,amsmath,amssymb]{revtex4}
\usepackage{booktabs}
\usepackage{mathrsfs}
\usepackage{epsfig}
\usepackage{graphicx}% Include figure files
\usepackage{dcolumn}% Align table columns on decimal point
\usepackage{bm}% bold math
\usepackage{amsmath}
\usepackage{slashed}       % slashed p
\usepackage{float}
\usepackage{natbib}
\usepackage[dvipdfm,  %pdflatex,pdftex    这里决定运行文件的方式不同
            pdfstartview=FitH,
            CJKbookmarks=true,
            bookmarksnumbered=true,
            bookmarksopen=true,
            colorlinks, %注释掉此项则交叉引用为彩色边框(将colorlinks和pdfborder 同时注释掉)
            pdfborder=001,   %注释掉此项则交叉引用为彩色边框
            linkcolor=red,
            anchorcolor=red,
            citecolor=blue
            ]{hyperref}
\usepackage{multirow}

\let\jnfont=\rm
\def\NPB#1,{{\jnfont Nucl.\ Phys.\ B }{\bf #1},}
\def\PLB#1,{{\jnfont Phys.\ Lett.\ B }{\bf #1},}
\def\EPJC#1,{{\jnfont Eur.\ Phys.\ Jour.\ C }{\bf #1},}
\def\PRD#1,{{\jnfont Phys.\ Rev.\ D }{\bf #1},}
\def\PRL#1,{{\jnfont Phys.\ Rev.\ Lett.\ }{\bf #1},}
\def\MPLA#1,{{\jnfont Mod.\ Phys.\ Lett.\ A }{\bf #1},}
\def\JPG#1,{{\jnfont J.\ Phys.\ G}{\bf #1},}
\def\CTP#1,{{\jnfont Commun.\ Theor.\ Phys.\ }{\bf #1},}
\def\ZPC#1,{{\jnfont Z.\ Phys.\ C }{\bf #1},}
\def\JHEP#1,{{\jnfont JHEP \ }{\bf #1},}
\def\lsim{\raise0.3ex\hbox{$<$\kern-0.75em\raise-1.1ex\hbox{$\sim$}}}
\def\gsim{\raise0.3ex\hbox{$>$\kern-0.75em\raise-1.1ex\hbox{$\sim$}}}

\begin{document}

\title{The properties of the Higgs bosons and Pair Production of the SM-like Higgs Boson in $\lambda$-SUSY at the LHC}

\author{Haijing Zhou$^1$, Zhaoxia Heng$^1$, Dongwei Li$^2$}
\affiliation{
\small  $^1$Department of Physics,
        Henan Normal University, Xinxiang 453007, China \\
\small  $^2$Department of Foundation,
        Henan Police College, Zhengzhou 450000, China
\vspace{1cm}}

\begin{abstract}
Compared with the MSSM or the NMSSM with a low $\lambda$, $\lambda$-SUSY theory with a large $\lambda$ around one has been deemed as a most natural realization of NMSSM. In this work, we treat the next-to-lightest CP-even Higgs boson as the SM-like Higgs boson in $\lambda$-SUSY and study the properties of the Higgs bosons and the pair production of the SM-like Higgs boson by considering various experiment constraints. We find that naturalness plays an important role in selecting the parameter space of $\lambda$-SUSY. In the most natural region of parameter space, the triple self coupling of the SM-like Higgs boson compared with its SM prediction may get enhanced by a factor about 7, and the most dominant contribution to the Higgs pair production comes from the triple self coupling of the SM-like Higgs boson and the production rate can be greatly enhanced, maximally 10 times larger than the SM prediction.
\end{abstract}

\pacs{14.80.Da,12.60.Jv}

\maketitle

\section{Introduction}

Since a near 125~GeV scalar particle was discovered in 2012, both the ATLAS and CMS collaborations have accumulated overwhelming evidence \cite{Aad:2012tfa,Chatrchyan:2012ufa,ATLAS:2013sla,CMS:2013yva} and the experimental data manifest that the properties of this new particle are roughly coincident with the Higgs boson predicted by the Standard Model (SM). However, it is yet unclear whether this particle is the Higgs boson in the SM. Moreover, the slight excess of the di-photon signal rate for the 125 GeV scalar reported by the LHC experiment \cite{ATLAS&CMS,ATLAS12} can be interpreted reasonably in new physics models, such as the Minimal Supersymmetric Standard Model (MSSM) \cite{Heinemeyer:2011aa,Draper:2011aa,Carena:2011aa,Cao:2011sn,Christensen:2012ei} and the Next-to-Minimal Supersymmetric Model (NMSSM) \cite{Ellwanger:2011aa,Gunion:2012zd,Cao:2012fz,Vasquez:2012hn,Benbrik:2012rm, Choi:2012he,King:2012tr,Badziak:2013bda,Moretti:2013lya,Ellwanger:2009dp,
Maniatis:2009re,Cao:2014kya}. Although the MSSM can accommodate a 125 GeV Higgs boson, it needs sizable radiative corrections from
top/stop loops, which will lead to some fine-tuning and seems unnatural. As the simplest realization of non-minimal extension of SUSY at the weak scale, NMSSM can offer additional contributions to the tree-level Higgs mass, which are not present in the MSSM, to accommodate a 125 GeV Higgs with significantly less fine-tuning.

Compared with the particle content of the MSSM, the NMSSM adds an additional gauge singlet superfield $\hat{S}$. Consequently, the form of its superpotential is given by\cite{Ellwanger:2009dp,Maniatis:2009re,Cao:2014kya}:
\begin{eqnarray}
W^{\rm NMSSM} & = & W_F + {\lambda}{\hat{H_u}}{\cdot}{\hat{H_d}}{\hat{S}}
+{1\over 3}{\kappa}{{\hat{S}}^3},
\label{superpotential}
\end{eqnarray}
with $ W_F $ denoting the superpotential of the MSSM without the ${\mu}$-term, and $\lambda$, $\kappa$ being the dimensionless parameters that describe the interactions among the superfields. When the singlet field $\hat{S}$ develops a vacuum expectation value $s$, an effective $\mu$ is generated with ${\mu}_{eff} \equiv {\lambda}{s}$. The inclusion of the $\hat{S}$ allows an additional contribution proportional to $\lambda^2$ to the Higgs potential, which will enhance the tree-level Higgs mass and reduce fine-tuning \cite{Kang:2012sy}.
Traditionally, many studies focused on  $\lambda \lesssim 0.7$, where the theory remains perturbative up to the grand unification scale ($10^{16}$~GeV).  In this scenario, the size of the additional contribution to the Higgs mass is restricted so that the issue of naturalness for 125~GeV Higgs is only partially addressed. In order to further reduce fine-tuning, the NMSSM with a relatively large $\lambda$ around one (dubbed as $\lambda-$SUSY) is emphasized
\cite{Barbieri:2006bg,Cao:2008un,Perelstein:2012qg,Hall:2011aa,Agashe:2012zq,Gherghetta:2012gb,Barbieri:2013hxa,Farina:2013fsa}.

Due to the peculiarity of $\lambda-$SUSY theory, it may easily elevate the tree-level mass of the SM-like Higgs boson (denoted by $h$ hereafter) larger than 125 GeV. So the structure of this boson must be a mixture of containing sizable singlet and/or non-SM doublet components \cite{Hall:2011aa,Agashe:2012zq,Gherghetta:2012gb}, which induces the couplings of SM-like Higgs boson may deviate remarkably from SM predictions. Moreover, the SM-like Higgs pair production in $\lambda-$SUSY may be enhanced significantly due to large trilinear self coupling of SM-like Higgs boson, which plays an indispensable role in constructing the Higgs potential.

Motivated by these arguments, we firstly study some features of the Higgs sector in $\lambda-$SUSY by considering various experimental constraints same as in work \cite{Cao:2014kya}, then we explore the properties of $h$ and the heaviest CP-even Higgs boson. Moreover, we also investigate the SM-like Higgs pair production and compare the results in $\lambda-$SUSY with the SM predictions. We ignore the lightest CP-even Higgs boson for its little contribution to the SM-like Higgs pair production because it mainly consists of the singlet field $S$ for the surviving samples in this work.

The paper is organized as follows. In section II, we describe the features of the Higgs sector in $\lambda-$SUSY. In section III, we scan the parameter space of $\lambda$-SUSY by considering various theoretical and experimental constraints. Then in the allowed parameter space, we study the properties of SM-like Higgs boson $h$ and the heaviest CP-even Higgs boson $h_3$, and also investigate the pair production rates of $h$ normalized to its SM prediction. Finally, in section IV, the conclusions are given.

\section{Higgs Sector in $\lambda$-SUSY}
In $\lambda$-SUSY theory, the scalar potential of the Higgs fields $H_u$, $H_d$ and $S$ consists of the contributions from the usual F-term and D-term, and also the soft breaking terms, which are given by:
\begin{eqnarray}
V_{soft}^{NMSSM} &=& {\tilde {m}}_u^2{|{H_u}|^2} + {\tilde{m}}_d^2{|{H_d}|^2} + {\tilde{m}}_S^2{|S|^2}
+({\lambda}{A_{\lambda}}{S}{H_u}{\cdot}{H_d} + {1\over 3}{\kappa}{A_{\kappa}}{\hat{S}}^3 + h.c.).
\label{soft-breaking}
\end{eqnarray}
Therefore, the Higgs sector Lagrangian includes 7 free parameters:
\begin{eqnarray}
p_i^{susy} &=& \{ {\lambda}, {\kappa}, {\tilde {m}}_u^2, {\tilde {m}}_d^2, {\tilde{m}}_S^2,
 {A_{\lambda}}, {A_{\kappa}} \}.
\label{parameters}
\end{eqnarray}

Like the general treatment of the multiple-Higgs theory, the Higgs fields of NMSSM can be written as follows:
\begin{eqnarray}
H_u =\left( \begin{array} {c} H_u^+ \\ {\upsilon}_u + {{\varphi_u+i\phi_u}\over \sqrt{2}} \end{array} \right),
~H_d=\left( \begin{array} {c} {\upsilon}_d+{{\varphi_d+i\phi_d} \over \sqrt{2}}\\H_d^-   \end{array} \right),
~S = s + {1\over \sqrt{2}} ({\sigma} + i{\xi})
\end{eqnarray}
with ${\upsilon}_u$, ${\upsilon}_d$ and $s$ representing the vacuum expectation values of the fields $H_{u}$, $H_{d}$ and $S$, respectively.
However, in order to clearly see the Higgs particle implication on the LHC results, one usually rewrite the Higgs fields with
one of them corresponds to the SM Higgs field \cite{higgsfield},
\begin{eqnarray}
H_1 =\left( \begin{array}{c}    H^+ \\ {S_1+iP_1} \over \sqrt{2}  \end{array} \right),
~H_2=\left( \begin{array}{c}   G^+ \\ {\upsilon} + {{S_2+iG^0} \over \sqrt{2}}  \end{array} \right),
~H_3 = s + {1\over \sqrt{2}} (S_3 + i{P_2}).
\label{higgs-filed}
\end{eqnarray}
where
\begin{eqnarray}
H_1 = \cos{\beta}H_{u} + {\varepsilon}\sin{\beta}H_{d}^*,
~H_2 = \sin{\beta}H_{u} - {\varepsilon}\cos{\beta}H_{d}^*,
~H_3 = S,
\end{eqnarray}
with ${\varepsilon}_{12} = -{\varepsilon}_{21} = 1$, ${\varepsilon}_{11} = {\varepsilon}_{22} = 0$, $\tan{\beta} \equiv  {\upsilon}_u / {\upsilon}_d$ and $\upsilon=\sqrt{{\upsilon}_u^2+{\upsilon}_d^2}$.
Eq.(\ref{higgs-filed}) indicates that the Higgs sector of the NMSSM contains three physical CP-even Higgs bosons formed by
the fields $S_1$, $S_2$ and $S_3$, two physical CP-odd Higgs bosons formed by the fields $P_1$ and $P_2$,
as well as one charged Higgs $H^+$.

Through the minimization conditions of the scalar potential, the soft breaking parameters ${\tilde {m}}_u^2$, ${\tilde {m}}_d^2$, ${\tilde{m}}_S^2$ in Eq.(\ref{parameters}) can be expressed in terms of $m_Z$, $\tan\beta$ and ${\mu}_{eff}$. Therefore, the
Higgs sector Lagrangian can also be described by the following six parameters,
\begin{eqnarray}
\lambda, ~\kappa, ~\tan\beta, ~\mu\equiv {\lambda}{s},~ M_A, M_P
\label{parameters2}
\end{eqnarray}
with $M_A^2=\frac{2\mu (A_\lambda +\kappa s)}{\sin2\beta}$ and
$M_P^2=\lambda^2\upsilon^2(\frac{M_A\sin2\beta}{2\mu})^2+\frac{3}{2}\lambda\kappa\upsilon^2\sin2\beta-3\kappa s A_\kappa$
representing the squared masses of the CP-odd fields $P_1$ and $P_2$, respectively. For the CP-even Higgs bosons in the basis($S_1$, $S_2$, $S_3$), the mass matrix is given by \cite {Ellwanger:2009dp,Maniatis:2009re}
\begin{eqnarray}
%\begin{align}
{\cal{M}}_{S,11}^2 &=& M^2_A+(m_Z^2-\lambda^2\upsilon^2)\sin^2{2\beta}, \nonumber\\
{\cal{M}}_{S,12}^2 &=& -{1\over 2} (m_Z^2-\lambda^2\upsilon^2)\sin{4\beta}, \nonumber\\
{\cal{M}}_{S,13}^2 &=& -\Big(\frac{M_A^2\sin2\beta}{2\mu}+\kappa\upsilon_s\Big)\lambda\upsilon \cos2\beta, \nonumber\\
{\cal{M}}_{S,22}^2 &=& m_Z^2\cos^2{2\beta}+\lambda^2\upsilon^2\sin^2{2\beta}, \nonumber\\
{\cal{M}}_{S,23}^2 &=& \Big[1-\Big(\frac{M_A\sin2\beta}{2\mu}\Big)^2-\frac{\kappa\sin2\beta}{2\lambda}\Big]2\lambda\mu\upsilon, \nonumber\\
{\cal{M}}_{S,33}^2 &=& \frac{1}{6}\lambda^2\upsilon^2\Big(\frac{M_A\sin2\beta}{\mu}\Big)^2+4(\kappa\upsilon_s)^2-\frac{1}{3}M^2_P.
\label{cp-even-higgs-mass}
%\end{align}
\end{eqnarray}

By diagonalizing the mass matrix for CP-even Higgs bosons in the basis ($S_1$, $S_2$, $S_3$), one can obtain the corresponding mass
eigenstates $h_i(i = 1, 2, 3)$:
\begin{eqnarray}
h_i = \sum^3_{j=1}{V_{ij}{S_j}},
\label{higgs-eigenstates}
\end{eqnarray}
with $V_{ij}$ denoting the rotation matrix and $m_{h_1} < m_{h_2} < m_{h_3}$. The state $h_i$ with $|V_{i2}|^2 > 0.5$ is called
the SM-like Higgs boson (labeled as $h$) and $h_i$ with $|V_{i1}|^2 > 0.5$ is called non-SM doublet Higgs boson.
Moreover, we define $\bar{S}_i = V_{i3}$, $\bar{D}_i = V_{i1}$, and they conform to the following sum rules
\begin{eqnarray}
\bar{D}_1^2 +\bar{D}_2^2 + \bar{D}_3^2 =1,
~{\bar{S}_1^2} +\bar{S}_2^2 + \bar{S}_3^2 =1.
\end{eqnarray}
Obviously, $|\bar{S}_i|^2$ represents the singlet component of the physical state $h_i$ and $|\bar{D}_i|^2$ represents
the non-SM doublet component of $h_i$.

In the NMSSM, without the mixing among the states $S_i$, the mass of the SM-like Higgs boson $h$ at tree level is given by
\begin{eqnarray}
m_{h,tree}^2 \simeq {m_{Z}^2}\cos^2{2\beta} + {\lambda}^2{\upsilon}^2{\sin^2{2\beta}},
\label{mh-tree}
\end{eqnarray}
where the second term on the right side is additional contribution originating from the coupling
$\lambda\hat{H_u}\cdot\hat{H_d}\hat{S}$ in the superpotential. Eq.(\ref{mh-tree}) indicates that
$m_{h,tree}$ can reach 125~GeV with a large $\lambda$ around one, which may be realized in the so-called
$\lambda$-SUSY theory \cite{Barbieri:2006bg,Cao:2008un}. $\lambda$-SUSY theory is only an effective lagrangian at the weak scale
and restricted to remain perturbative up to about 10 TeV, which renders the parameters $\lambda$ and $\kappa$ at weak scale
satisfying the following relation \cite{Barbieri:2006bg}
\begin{eqnarray}
0.17{\lambda}^2 + 0.26{\kappa}^2 \lesssim 1.
\label{lam-kap}
\end{eqnarray}

In order to measure the naturalness of the $\lambda$-SUSY theory, two fine tuning quantities are defined as follows \cite{Barbieri:2006bg,Ellwanger:2011mu}:
\begin{eqnarray}
{\Delta}_Z = \underset{i}{max}{|{\partial{\log{m_Z^2}} \over \partial{\log{p_i}}}|},
~{\Delta}_h = \underset{i}{max}{|{\partial{\log{m_h^2}} \over \partial{\log{p_i}}}|},
\end{eqnarray}
where $p_i$ includes the SUSY parameters at the weak scale listed in Eq.(\ref{parameters}) and also top quark Yukawa coupling $Y_t$.
We adopt the formulae presented in \cite{Ellwanger:2011mu} and \cite{Barbieri:2006bg} to calculate ${\Delta}_Z$ and ${\Delta}_h$.
Obviously, with smaller values of ${\Delta}_Z$ and ${\Delta}_h$, the $\lambda$-SUSY theory is more natural in predicting
$m_Z$ and $m_h$. Therefore, we use ${\Delta}_Z$ and ${\Delta}_h$ to estimate the goodness of the surviving samples
and take $\max$$\{$ ${\Delta}_Z, {\Delta}_h \} \le 50$ as a criterion for naturalness.

In the following discussions, we only consider the scenario in which the next-to-lightest CP-even Higgs boson is
SM-like Higgs boson $h$, and compare the results with work \cite{Cao:2014kya}, which takes the lightest CP-even
Higgs boson as the SM-like Higgs boson.

\section{Numerical results and discussions}
In this work we use the package NMSSMTools-4.0.0 to scan over the parameter space of $\lambda$-SUSY by considering
various experimental and theoretical constraints,
\begin{eqnarray}
& 0.7 < {\lambda} \le 2,~~ 0 < \kappa \le 2, ~~100~\rm GeV < M_A, M_P,  \mu \le 3~TeV,& \nonumber\\
& 100~\rm GeV \le  M_{Q_3}, M_{U_3}  \le  2~TeV, ~~|A_t|  \le   5~TeV,& \nonumber\\
& 100~\rm GeV \le m_{\tilde{l}}, M_2 \le 1~TeV,~~ 1 \le \tan{\beta} \le 15,&
\label{parameter-regions}
\end{eqnarray}
with all the parameters defined at the 1TeV scale. Most of the constraints are implemented in the package NMSSMTools.
Furthermore, we take 120 GeV $\leq m_h \leq$ 130 GeV and consider the indirect constraints from the electroweak precision data, which strongly affect $\tan\beta$ in $\lambda$-SUSY.

For the surviving samples, we also perform a fit to the Higgs data from ATLAS \cite{ATLAS2014}, CMS \cite{CMS2014} and CDF+D0 \cite{Aaltonen:2013ioz},
and adopt the method introduced in \cite{Espinosa:2012ir,Giardino:2012ww} to calculate corresponding $\chi^2$. We obtain $\chi^2_{min}$ =11.7 with Higgs
data in 2014. In the following discussions, we consider three types of surviving samples and focus on samples
with ${\chi}^2 \le 25$, which corresponds to the samples consistent with the Higgs data at 95\% C.L..
\begin{itemize}
  \item[${\bullet}$] Type-I samples: samples satisfying both $\chi^2 \le 25$ and $\max$$\{$ ${\Delta}_Z, {\Delta}_h \} \le 50$,
  which are regarded as the physical samples in our discussion.
  \item[${\bullet}$] Type-II samples: samples satisfying both $\chi^2 \le 25$ and $\max$$\{$ ${\Delta}_Z, {\Delta}_h \} > 50$,
  which coincide with the experiments but are not favored by the fine tuning argument.
  \item[${\bullet}$] Type-III samples: samples satisfying $\chi^2 > 25$, which are of less interest than the previous two types.
\end{itemize}

\begin{figure}
\includegraphics[width=15cm]{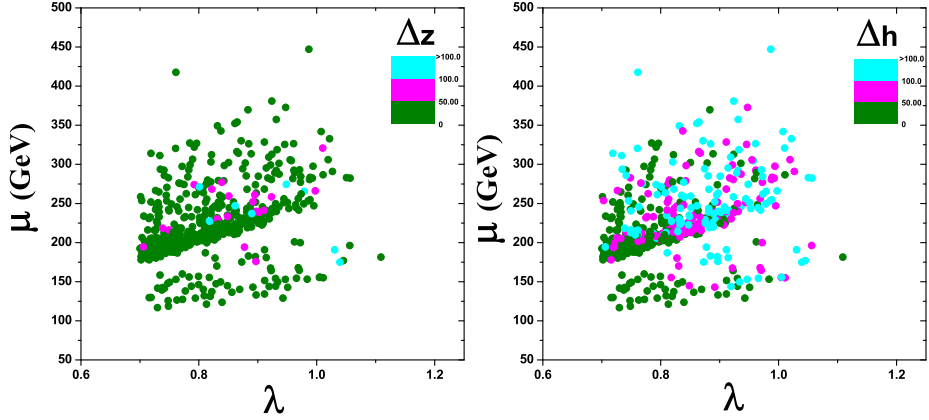}
\vspace{-0.5cm} \caption{Surviving samples satisfying ${\chi}^2 \le 25$, projected on the plane of $\mu$ versus $\lambda$. For these samples, their corresponding values of ${\Delta}_Z$ and ${\Delta}_h $ are displayed with different colors.}
\label{lam-mu}
\end{figure}

Since the $\lambda$-SUSY theory is more natural in predicting $m_Z$ and $m_h$ than the MSSM, we display the characters
of ${\Delta}_Z$ and ${\Delta}_h $ in $\lambda$-SUSY. In fig.\ref{lam-mu} we show the surviving samples on the plane of
$\mu$ versus $\lambda$ with the corresponding values of ${\Delta}_Z$ and ${\Delta}_h $ in different colors.
The figure manifests that the largest value of $\lambda$ may reach to 1.1, which is different from the scenario with
the lightest CP-even Higgs boson as the SM-like Higgs boson, in which the value of $\lambda$ can reach
to 1.8 \cite{Cao:2014kya}. This is because the mixing of the fields $S_2$ and $S_3$ can push up the SM-like Higgs boson mass
when the next-to-lightest CP-even Higgs boson is SM-like.

%%%%%%%%%%%%%%%%%%%%%%%%%%%%%%%%%%%%%%%%%%%%%%%%%%%%%%%%%%%%%%%%%%%%%%%%%%%%%%%%%[width=19cm]
%%%%%%%%%%%%%%%%%%%%%%%%%%%%%%%%%%%%%%%%%%%%%%%%%%%%%%%%%%%%%%%%%%%%%%%%%%%%%%%%%%%%
%%%%%%%%%%%%%%%%%%%%%%%%%%%%%%%%%%%%%%%%%%%%%%%%%%%%%%%%%%%%%%%%%%%%%%%%%%%%%%%%%[width=19cm]
\begin{figure}
\includegraphics[width=15cm]{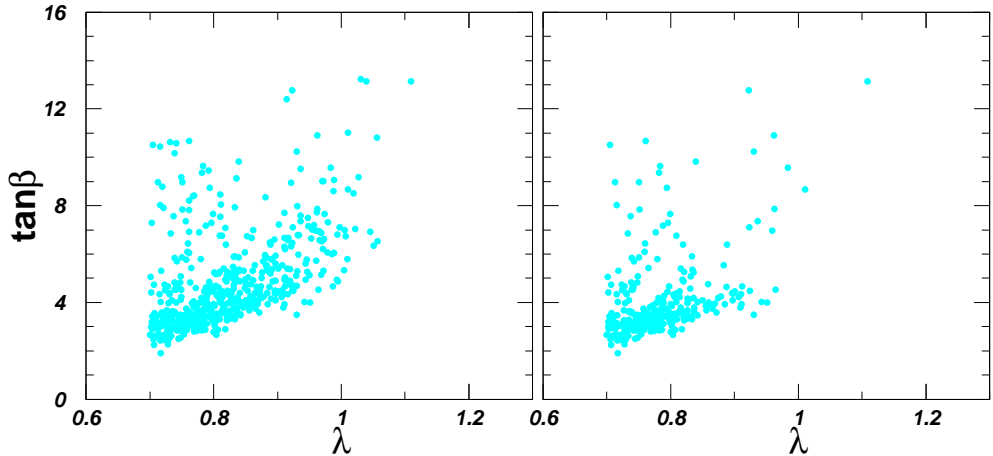}
\vspace{-0.5cm} \caption{ Surviving samples satisfying ${\chi}^2 \le 25$, projected on the plane of $\tan{\beta}$ versus $\lambda$.
Samples in the right panel are further required to satisfy $\max$ $\{$ ${\Delta}_Z, {\Delta}_h \} \le 50$.}
\label{lam-tanb}
\end{figure}
%%%%%%%%%%%%%%%%%%%%%%%%%%%%%%%%%%%%%%%%%%%%%%%%%%%%%%%%%%%%%%%%%%%%%%%%%%%%%%%%%%%%

Due to the important role of $\tan{\beta}$ and $\lambda$ in $\lambda$-SUSY, we show the correlation between them in Fig.\ref{lam-tanb}.
The surviving samples satisfy ${\chi}^2 \le 25$ and those in the right panel are further required to satisfy
$\max$ $\{$ ${\Delta}_Z, {\Delta}_h \} \le 50$. The figure indicates that $\tan\beta$ tends to increase with the increase
of $\lambda$ and some samples with relatively small values of $\tan\beta$ are excluded after requiring
$\max$ $\{$ ${\Delta}_Z, {\Delta}_h \} \le 50$. This is because a large value of $\tan\beta$ can reduce the tree level
Higgs boson mass $m_{h,tree}$, which is able to cancel out the mixing effect of the fields $S_2$ and $S_3$ in order to obtain a 125 GeV Higgs boson.

In the following discussions, we investigate the properties of the next-to-lightest and heaviest CP-even Higgs bosons for the above three types of samples. We pay particular attention to the features of these bosons that differentiate from \cite{Cao:2014kya}.

\subsection{Properties of the Next-to-Lightest CP-even Higgs Boson $h_2$}
Throughout this work we take the next-to-lightest CP-even Higgs boson $h_2$ as the SM-like Higgs boson $h$, so
in the following discussions, we explore the features of this boson and also its coupling information.
%%%%%%%%%%%%%%%%%%%%%%%%%%%%%%%%%%%%%%%%%%%%%%%%%%%%%%%%%%%%%%%%%%%%%%%%%%%%%%%%%[width=19cm]
\begin{figure}[!htb]
\includegraphics[width=13cm]{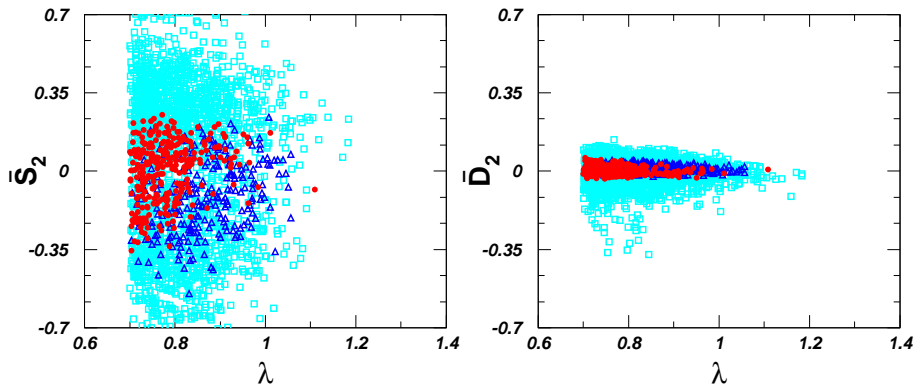}
\vspace{-0.5cm} \caption{Singlet component coefficient $\bar{S}_2$ and non-SM doublet component coefficient $\bar{D}_2$ of the SM-like Higgs boson as a function of $\lambda$ for Type-I sample (red bullet), Type-II sample (blue triangle) and Type-III sample
( sky-blue square).}
\label{hsmprop}
\end{figure}
%%%%%%%%%%%%%%%%%%%%%%%%%%%%%%%%%%%%%%%%%%%%%%%%%%%%%%%%%%%%%%%%%%%%%%%%%%%%%%%%%%%%
%%%%%%%%%%%%%%%%%%%%%%%%%%%%%%%%%%%%%%%%%%%%%%%%%%%%%%%%%%%%%%%%%%%%%%%%%%%%%%%%%[width=19cm]
\begin{figure}[!htb]
\includegraphics[width=11cm]{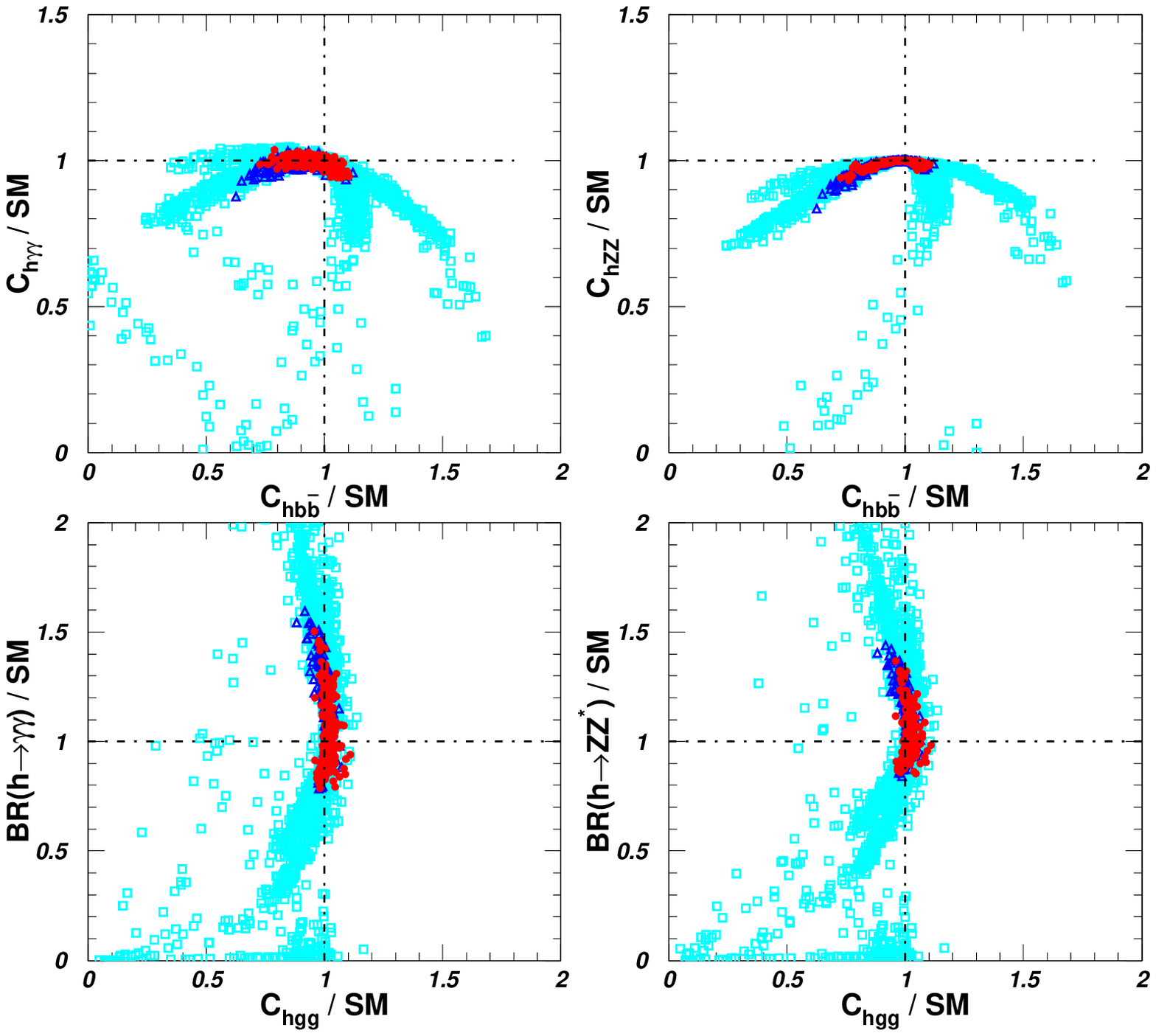}
\includegraphics[width=11.5cm]{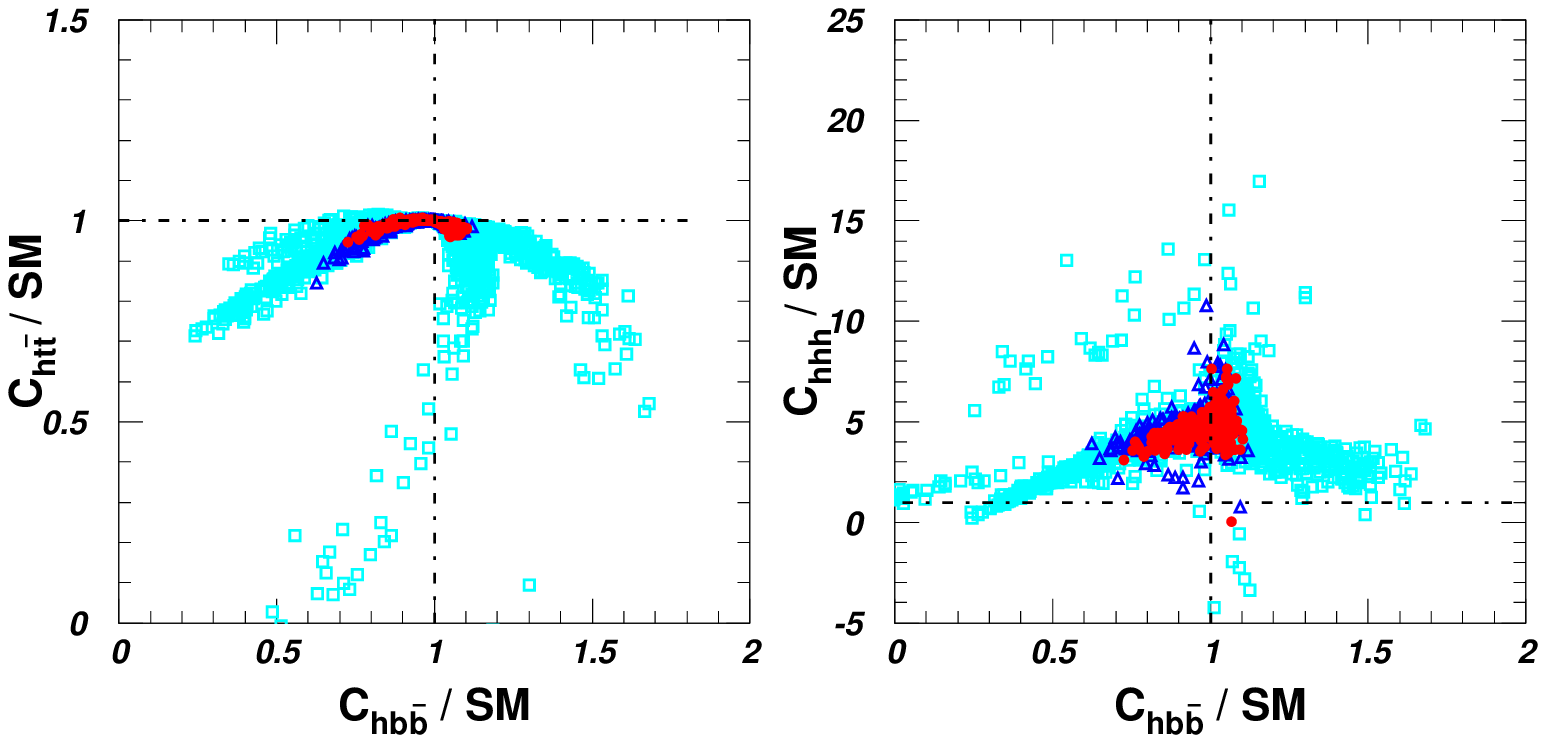}
\vspace{-0.5cm} \caption{Same as Fig.\ref{hsmprop}, but showing the coupling information of the SM-like Higgs boson $h$.}
\label{h-gzbth}
\end{figure}
%%%%%%%%%%%%%%%%%%%%%%%%%%%%%%%%%%%%%%%%%%%%%%%%%%%%%%%%%%%%%%%%%%%%%%%%%%%%%%%%%%%%

In Fig.\ref{hsmprop}, we project the three types of surviving samples on the plane of $\bar{S}_2-\lambda$ and $\bar{D}_2-\lambda$,
where $\bar{S}_2$ and $\bar{D}_2$ denote the singlet component coefficient and non-SM doublet component coefficient of $h$ respectively.
The figure shows that the range of the values of $|\bar{S}_2|$ is always larger than that of $|\bar{D}_2|$.
As is showed in Fig.\ref{hsmprop}, $|\bar{S}_2|$ may exceed 0.7 and be smaller than 0.35 before and after considering the Higgs data at $95\%$ C.L. respectively. In comparison, $|\bar{D}_2|$ reaches maximally about 0.35, and it is less than 0.1 with the Higgs data
at $95\%$ C.L.. This is because the constraints we considered have put weak restrictions on the element ${\cal{M}}_{S,33}^2$ of the CP-even Higgs mass matrix due to the singlet nature of the field $S_3$. Compared with the Figure 3 in \cite{Cao:2014kya}, we find the values of
the singlet component coefficient $|\bar{S}_2|$ has a wider range in our discussion than in the work \cite{Cao:2014kya}, which take the lightest CP-even Higgs boson as the SM-like Higgs boson. The reason is that $m_{h,tree}$ will be easily lifted to much larger than 125 GeV and the sizable singlet component can effectively pull down the mass.

In Fig.\ref{h-gzbth}, we exhibit the coupling information of $h$. This figure indicates that, after considering the Higgs data at $95\%$ C.L., the normalized couplings $C_{h\gamma\gamma}/SM$, $C_{hZZ}/SM$, $C_{hgg}/SM$ and $C_{h\bar{t}t}/SM$ are limited within $20\%$ deviation from unity, and the couplings $C_{h\bar{b}b}/SM$ are allowed to vary in relatively wider ranges, at $40\%$ deviating from unity. The normalized branching ratios $Br(h \to \gamma\gamma )/SM$ and $Br(h \to ZZ )/SM$ may vary from 0.8 to 1.5. For the Higgs triple self coupling $C_{hhh}/SM$, the right panel of the last row in Fig.\ref{h-gzbth} indicates that it can only reach 7 and 10 for the Type-I samples and Type-II samples respectively.

\subsection{Properties of the heaviest CP-even Higgs Boson $h_3$}
%%%%%%%%%%%%%%%%%%%%%%%%%%%%%%%%%%%%%%%%%%%%%%%%%%%%%%%%%%%%%%%%%%%%%%%%%%%%%%%%%[width=19cm]
\begin{figure}[!htb]
\includegraphics[width=17cm]{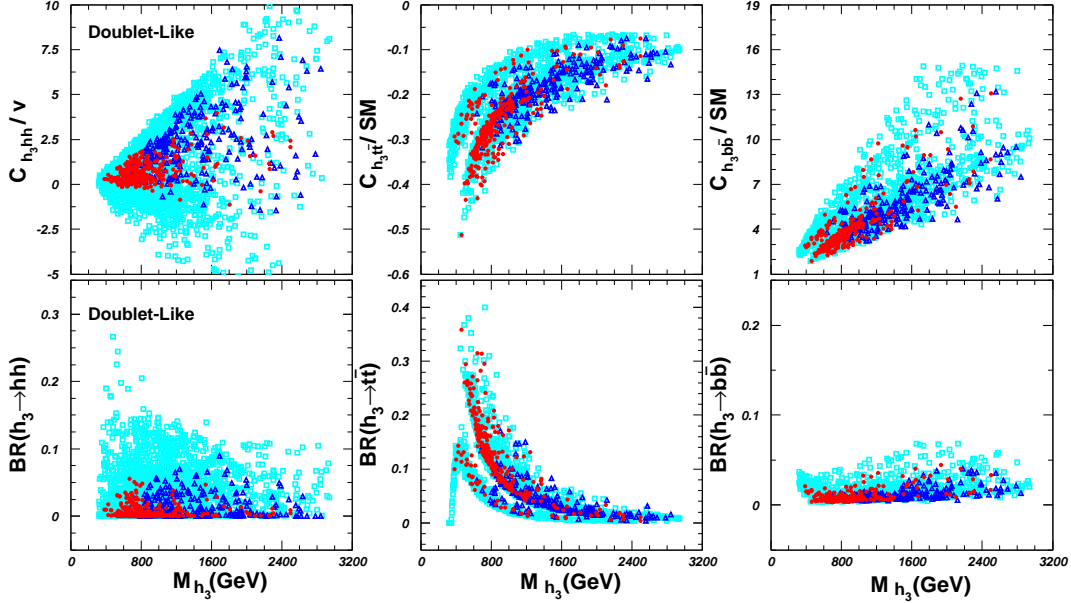}
\vspace{-0.5cm} \caption{Same as Fig.\ref{hsmprop}, but showing the couplings and branching ratios of doublet dominated ${h_3}$ as a function of $M_{h_3}$.}
\label{h3br_d}
\end{figure}
%%%%%%%%%%%%%%%%%%%%%%%%%%%%%%%%%%%%%%%%%%%%%%%%%%%%%%%%%%%%%%%%%%%%%%%%%%%%%%%%%%%%
We here simply study the properties of the heaviest CP-even Higgs boson $h_3$. We find that the non-SM doublet component coefficient $\bar{D}_3^2$ of $h_3$ is over 0.9 for Type-I and Type-II samples, that is to say, ${h_3}$ is dominated by the non-SM doublet,
which can be foreordained.

In Fig.\ref{h3br_d}, we show the normalized couplings of the ${h_3}$ such as $C_{{h_3}hh}/{\upsilon}$, $C_{{h_3}\bar{t}t}/SM$ and $C_{{h_3}\bar{b}b}/SM$ as functions of $M_{h_3}$ and also plot the branching ratios of $h_3 \to hh$, $h_3 \to \bar{t}t$ and $h_3 \to \bar{b}b$. We can learn the following features from Fig.\ref{h3br_d}: (1) For all the surviving samples, $m_{h_3} \gtrsim$ 400 GeV; (2) The normalized coupling $C_{{h_3}hh}/{\upsilon}$ may still be large with the maximum value reaching 5, and $C_{{h_3}\bar{b}b}/SM$ is larger than 1 with maximum value of 10 in optimal case. While for the normalized coupling $C_{{h_3}\bar{t}t}/SM$, it is smaller than 1 and $0.2\le \big|C_{{h_3}\bar{t}t}/SM\big|\le 0.4$. (3) For 500 GeV $\le m_{h_3} \le$ 1000 GeV, $h_3 \to \bar{t}t$ may  act as the dominant decay channel of $h_3$.

%%%%%%%%%%%%%%%%%%%%%%%%%%%%%%%%%%%%%%%%%%%%%%%%%%%%%%%%%%%%%%%%%%%%%%%%%%%%%%%%%[width=19cm]
\begin{figure}[!htb]
\includegraphics[width=15cm]{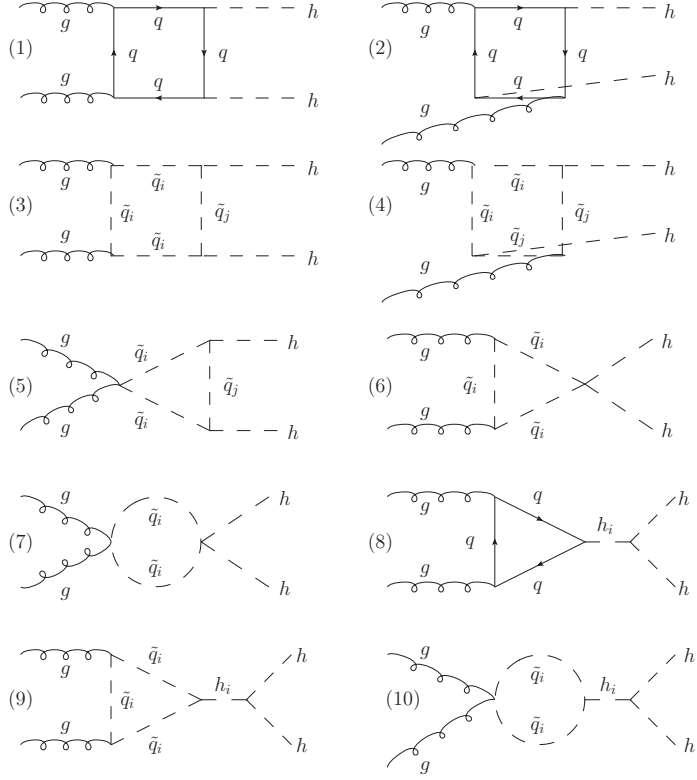}
\vspace{-0.5cm} \caption{Feynman diagrams for the pair production of the SM-like Higgs boson via gluon fusion in $\lambda$-SUSY with $h_i$ denoting a CP-even Higgs (i = 1, 2, 3) and $\tilde{q}_{i,j}$(i, j = 1, 2) denoting a squark.
The diagrams with initial gluons or final Higgs bosons interchanged are not shown here.}
\label{Feynman-diagram}
\end{figure}
%%%%%%%%%%%%%%%%%%%%%%%%%%%%%%%%%%%%%%%%%%%%%%%%%%%%%%%%%%%%%%%%%%%%%%%%%%%%%%%%%%%%
\subsection{Higgs Pair Production at the LHC}
As a rare process at the LHC compared with other Higgs production channels, the Higgs pair production plays a significant role in extracting Higgs self interaction $C_{hhh}$, which is indispensable to reconstruct the Higgs potential and finally
interpret the mechanism of the electroweak symmetry breaking \cite{Dolan:2012rv,McCullough:2013rea,Wu:2015nba,Cao:2015oaa}. So the studies on the Higgs pair production in $\lambda$-SUSY should be carefully investigated.

In SUSY, the Higgs pair production may proceed through the diagrams (1)-(10) in
Fig.\ref{Feynman-diagram} \cite{Nhung:2013lpa,Ellwanger:2013ova,Han:2013sga}, where
the diagrams with initial gluons or final Higgs bosons interchanged are not shown, and we only consider the contributions from the third generation quarks and squarks due to their large Yukawa couplings. The SUSY prediction on the production rate may significantly deviate from the SM prediction because the contribution to the amplitude from SUSY is of the same perturbation order as that from the SM. Based on previous studies in the SUSY \cite{Liu:2013woa,No:2013wsa,Barger:2014taa,Cao:2013si,Cao:2013cfa}, we learn that the Higgs pair production rate may significantly enhanced mainly through the following three mechanisms: (i)Through the loops mediated by stops \cite{Cao:2013si}.
The major contributions come from diagrams (3)-(5) of Fig.\ref{Feynman-diagram}, and the quantitative amplitude can be given by
${\cal{M}} \sim \alpha^2_s Y_t^2
\Big( c_1sin^22\theta_t \frac{A^2_t}{m^2_{\tilde{t}_1}}
+c_2 \frac{A^2_t}{m^2_{\tilde{t}_2}} \Big)$ with $\theta_t$ being the mixing angle of stops and $c_1, c_2$ denoting dimensionless coefficients; (ii)Through large Higgs self coupling \cite{Cao:2013cfa}; (iii)Through the resonant effect of $h_i$ \cite{Liu:2013woa,No:2013wsa,Barger:2014taa}. In this work, only the heaviest CP-even Higgs $h_3$ can be on-shell produced by gg or $b{\bar b}$ initial state.

%%%%%%%%%%%%%%%%%%%%%%%%%%%%%%%%%%%%%%%%%%%%%%%%%%%%%%%%%%%%%%%%%%%%%%%%%%%%%%%%%[width=19cm]
\begin{figure}[!htb]
\includegraphics[width=17cm]{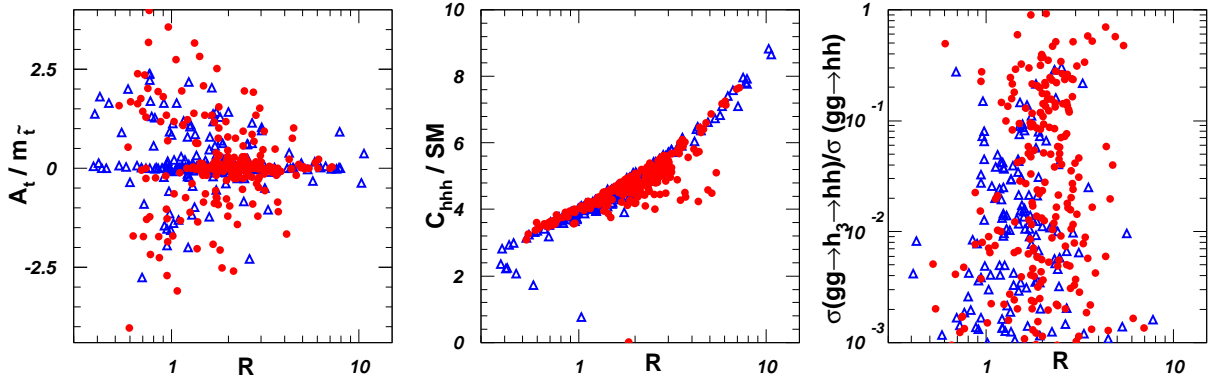}
\vspace{-0.5cm} \caption{The scatter plots of the Type-I samples (red bullet) and Type-II samples (blue triangle), showing $A_t/ m_{\tilde{t}}$, $C_{hhh} / SM$ and the pure resonant s-channel contribution to the pair production as a function of $R$ in left panel, middle panel and right panel respectively.}
\label{atmst-hhh-h3hh}
\end{figure}
%%%%%%%%%%%%%%%%%%%%%%%%%%%%%%%%%%%%%%%%%%%%%%%%%%%%%%%%%%%%%%%%%%%%%%%%%%%%%%%%%%%%
In order to investigate which mechanism is to manily enhance the Higgs pair production rate, in fig.\ref{atmst-hhh-h3hh} we present $A_t / m_{\tilde{t}}$, $C_{hhh} / SM$, the pure resonant s-channel contribution to the pair production as a function of the normalized Higgs pair production rate R, where $R=\sigma(pp\to hh)/(\sigma^{LO}_{SM}(PP\to hh)|_{m_h=125~GeV})\simeq \sigma(pp\to hh)/(19fb)$. The figure shows that the Higgs pair production rate can be enhanced by about 10 times maximally and the main contributions come from the Higgs self coupling. In comparison with the scenario of the lightest Higgs boson as SM-like Higgs in work \cite{Cao:2014kya}, the enhancement of the Higgs pair production rate is unconspicuous. The main reason is, as the right panel of the last row in Fig.4 shows, that the enhancement of the Higgs triple self coupling $C_{hhh}/SM$ are less significant than that in \cite{Cao:2014kya}. In addition, the maximum value of the normalized Higgs pair production rate is roughly same with the case in NMSSM with $\lambda <$ 0.7 \cite{Cao:2013si}, where the enhancement is manily through the loops mediated by stops.

\section{Conclusion}\label{Sum}
In this work,  we investigate the properties of the SM-like Higgs boson in $\lambda$-SUSY theory, which corresponds to the NMSSM with a large $\lambda$ around one. Throughout this work, we treat the next-to-lightest CP-even Higgs boson as the SM-like Higgs boson. To make our study realistic, we firstly define quantities $\Delta_Z$ and $\Delta_h$ to measure the naturalness of the parameter points and consider the Higgs data at 95\% C.L.. Then we implement various experimental constraints on the parameter space of $\lambda$-SUSY. In the allowed parameter space, we investigate features of the SM-like Higgs boson and the heaviest CP-even Higgs boson, and also study the pair production of the SM-like Higgs boson.
As is shown in this work, we find the following features:
\begin{itemize}
   \item[$\bullet$] Considering the naturalness argument, $\lambda$ is less than 1.1. However, $\lambda$ can reach 1.8 in work \cite{Cao:2014kya}, which treats the lightest CP-even Higgs boson as the SM-like Higgs boson. Moreover, after considering the naturalness argument, the surviving samples are fewer, especially $\Delta_h$ plays more important role in selecting the parameter space of $\lambda$-SUSY.

  \item[$\bullet$] Same as the current conclusion in \cite{Cao:2014kya}, Higgs data at 95\% C.L. still allow for a sizable singlet component in $h$, which at most reaches $35\%$, while the non-SM doublet component is forbidden to be larger than $10\%$.
  \item[$\bullet$] The strength of the triple self coupling of $h$ can only reach 7 and 10 for the Type-I samples and Type-II samples respectively, which is smaller than the maximum value in work \cite{Cao:2014kya}.
  \item[$\bullet$] The heaviest CP-even Higgs boson $h_3$ is highly non-SM doublet dominated and its mass may be as light as 400 GeV. For 500 GeV $\le m_{h_3} \le$ 1000 GeV, $h_3 \to \bar{t}t$ may  act as the dominant decay channel.
  \item[$\bullet$] In most cases the dominant contribution to the Higgs pair production comes from the triple self coupling of the Higgs boson and the production rate can be greatly enhanced, maximally 10 times larger than the SM prediction.

\end{itemize}

\section*{Acknowledgement}
We thank Prof. Junjie Cao for helpful discussions. This work was supported in part by the National Natural Science
Foundation of China (NNSFC) under grant No. 11305050, and
by Specialized Research Fund for the Doctoral Program of Higher
Education with grant No. 20124104120001.

%%%%%%%%%%%%%%%%%%%%%%%%%%%%%%%%%%%%%%%%%%%%%%%%%%%%%%%%%%%%%%%%%%%%%%%%%%%%%%


\begin{thebibliography}{32}

%\cite{Aad:2012tfa}
\bibitem{Aad:2012tfa}
  G.~Aad {\it et al.}  [ATLAS Collaboration],
  %``Observation of a new particle in the search for the Standard Model Higgs boson with the ATLAS detector at the LHC,''
  Phys.\ Lett.\ B {\bf 716}, 1 (2012)
  [arXiv:1207.7214 [hep-ex]].
  %%CITATION = ARXIV:1207.7214;%%
  %4491 citations counted in INSPIRE as of 14 juin 2015

%\cite{Chatrchyan:2012ufa}
\bibitem{Chatrchyan:2012ufa}
  S.~Chatrchyan {\it et al.}  [CMS Collaboration],
  %``Observation of a new boson at a mass of 125 GeV with the CMS experiment at the LHC,''
  Phys.\ Lett.\ B {\bf 716}, 30 (2012)
  [arXiv:1207.7235 [hep-ex]].
  %%CITATION = ARXIV:1207.7235;%%
  %4409 citations counted in INSPIRE as of 14 Jun 2015

%\cite{ATLAS:2013sla}
\bibitem{ATLAS:2013sla}
  [ATLAS Collaboration],
  %``Combined coupling measurements of the Higgs-like boson with the ATLAS detector using up to 25 fb$^{-1}$ of proton-proton collision data,''
  ATLAS-CONF-2013-034, ATLAS-COM-CONF-2013-035.
  %%CITATION = ATLAS-CONF-2013-034, ATLAS-COM-CONF-2013-035;%%
  %280 citations counted in INSPIRE as of 14 Jun 2015

  %\cite{CMS:yva}
\bibitem{CMS:2013yva}
  [CMS Collaboration],
  %``Combination of standard model Higgs boson searches and measurements of the properties of the new boson with a mass near 125 GeV,''
  CMS-PAS-HIG-13-005.
  %%CITATION = CMS-PAS-HIG-13-005;%%
  %392 citations counted in INSPIRE as of 14 juin 2015

\bibitem{ATLAS&CMS}
  G. Aad et al.,
  [ATLAS Collaboration], Phys.\ Lett.\ B {\bf 716}, 1 (2012);
  S. Chatrchyan et al.,
  [CMS Collaboration], Phys.\ Lett.\ B {\bf 716}, 30 (2012).

\bibitem{ATLAS12}
 The ATLAS Collaboration ATLAS-CONF-2012-170;
 The CMS Collaboration CMS-PAS-HIG-12-045.

  %\cite{Heinemeyer:2011aa}
\bibitem{Heinemeyer:2011aa}
  S.~Heinemeyer, O.~Stal and G.~Weiglein,
  %``Interpreting the LHC Higgs Search Results in the MSSM,''
  Phys.\ Lett.\ B {\bf 710}, 201 (2012)
  [arXiv:1112.3026 [hep-ph]].
  %%CITATION = ARXIV:1112.3026;%%
  %236 citations counted in INSPIRE as of 24 Jun 2015

  %\cite{Draper:2011aa}
\bibitem{Draper:2011aa}
  P.~Draper, P.~Meade, M.~Reece and D.~Shih,
  %``Implications of a 125 GeV Higgs for the MSSM and Low-Scale SUSY Breaking,''
  Phys.\ Rev.\ D {\bf 85}, 095007 (2012)
  [arXiv:1112.3068 [hep-ph]].
  %%CITATION = ARXIV:1112.3068;%%
  %237 citations counted in INSPIRE as of 24 Jun 2015

  %\cite{Carena:2011aa}
\bibitem{Carena:2011aa}
  M.~Carena, S.~Gori, N.~R.~Shah and C.~E.~M.~Wagner,
  %``A 125 GeV SM-like Higgs in the MSSM and the $\gamma \gamma$ rate,''
  JHEP {\bf 1203}, 014 (2012)
  [arXiv:1112.3336 [hep-ph]].
  %%CITATION = ARXIV:1112.3336;%%
  %318 citations counted in INSPIRE as of 24 juin 2015

  %\cite{Cao:2011sn}
\bibitem{Cao:2011sn}
  J.~Cao, Z.~Heng, D.~Li and J.~M.~Yang,
  %``Current experimental constraints on the lightest Higgs boson mass in the constrained MSSM,''
  Phys.\ Lett.\ B {\bf 710}, 665 (2012)
  [arXiv:1112.4391 [hep-ph]].
  %%CITATION = ARXIV:1112.4391;%%
  %126 citations counted in INSPIRE as of 24 Jun 2015

  %\cite{Christensen:2012ei}
\bibitem{Christensen:2012ei}
  N.~D.~Christensen, T.~Han and S.~Su,
  %``MSSM Higgs Bosons at The LHC,''
  Phys.\ Rev.\ D {\bf 85}, 115018 (2012)
  [arXiv:1203.3207 [hep-ph]].
  %%CITATION = ARXIV:1203.3207;%%
  %111 citations counted in INSPIRE as of 24 juin 2015

  %\cite{Ellwanger:2011aa}
\bibitem{Ellwanger:2011aa}
  U.~Ellwanger,
  %``A Higgs boson near 125 GeV with enhanced di-photon signal in the NMSSM,''
  JHEP {\bf 1203}, 044 (2012)
  [arXiv:1112.3548 [hep-ph]].
  %%CITATION = ARXIV:1112.3548;%%
  %201 citations counted in INSPIRE as of 24 juin 2015

  %\cite{Gunion:2012zd}
\bibitem{Gunion:2012zd}
  J.~F.~Gunion, Y.~Jiang and S.~Kraml,
  %``The Constrained NMSSM and Higgs near 125 GeV,''
  Phys.\ Lett.\ B {\bf 710}, 454 (2012)
  [arXiv:1201.0982 [hep-ph]].
  %%CITATION = ARXIV:1201.0982;%%
  %121 citations counted in INSPIRE as of 24 juin 2015

  %\cite{Cao:2012fz}
\bibitem{Cao:2012fz}
  J.~J.~Cao, Z.~X.~Heng, J.~M.~Yang, Y.~M.~Zhang and J.~Y.~Zhu,
  %``A SM-like Higgs near 125 GeV in low energy SUSY: a comparative study for MSSM and NMSSM,''
  JHEP {\bf 1203}, 086 (2012)
  [arXiv:1202.5821 [hep-ph]].
  %%CITATION = ARXIV:1202.5821;%%
  %247 citations counted in INSPIRE as of 24 Jun 2015

  %\cite{Vasquez:2012hn}
\bibitem{Vasquez:2012hn}
  D.~A.~Vasquez, G.~Belanger, C.~Boehm, J.~Da Silva, P.~Richardson and C.~Wymant,
  %``The 125 GeV Higgs in the NMSSM in light of LHC results and astrophysics constraints,''
  Phys.\ Rev.\ D {\bf 86}, 035023 (2012)
  [arXiv:1203.3446 [hep-ph]].
  %%CITATION = ARXIV:1203.3446;%%
  %80 citations counted in INSPIRE as of 24 juin 2015

  %\cite{Benbrik:2012rm}
\bibitem{Benbrik:2012rm}
  R.~Benbrik, M.~Gomez Bock, S.~Heinemeyer, O.~Stal, G.~Weiglein and L.~Zeune,
  %``Confronting the MSSM and the NMSSM with the Discovery of a Signal in the two Photon Channel at the LHC,''
  Eur.\ Phys.\ J.\ C {\bf 72}, 2171 (2012)
  [arXiv:1207.1096 [hep-ph]].
  %%CITATION = ARXIV:1207.1096;%%
  %103 citations counted in INSPIRE as of 24 juin 2015


  %\cite{Choi:2012he}
\bibitem{Choi:2012he}
  K.~Choi, S.~H.~Im, K.~S.~Jeong and M.~Yamaguchi,
  %``Higgs mixing and diphoton rate enhancement in NMSSM models,''
  JHEP {\bf 1302}, 090 (2013)
  [arXiv:1211.0875 [hep-ph]].
  %%CITATION = ARXIV:1211.0875;%%
  %45 citations counted in INSPIRE as of 24 Jun 2015

  %\cite{King:2012tr}
\bibitem{King:2012tr}
  S.~F.~King, M.~Mühlleitner, R.~Nevzorov and K.~Walz,
  %``Natural NMSSM Higgs Bosons,''
  Nucl.\ Phys.\ B {\bf 870}, 323 (2013)
  [arXiv:1211.5074 [hep-ph]].
  %%CITATION = ARXIV:1211.5074;%%
  %76 citations counted in INSPIRE as of 24 Jun 2015

  %\cite{Badziak:2013bda}
\bibitem{Badziak:2013bda}
  M.~Badziak, M.~Olechowski and S.~Pokorski,
  %``New Regions in the NMSSM with a 125 GeV Higgs,''
  JHEP {\bf 1306}, 043 (2013)
  [arXiv:1304.5437 [hep-ph]].
  %%CITATION = ARXIV:1304.5437;%%
  %43 citations counted in INSPIRE as of 24 Jun 2015
  %\cite{Moretti:2013lya}

\bibitem{Moretti:2013lya}
  S.~Moretti, S.~Munir and P.~Poulose,
  %``125 GeV Higgs Boson signal within the complex NMSSM,''
  Phys.\ Rev.\ D {\bf 89}, no. 1, 015022 (2014)
  [arXiv:1305.0166 [hep-ph]].
  %%CITATION = ARXIV:1305.0166;%%
  %17 citations counted in INSPIRE as of 24 Jun 2015

%\cite{Ellwanger:2009dp}
\bibitem{Ellwanger:2009dp}
  U.~Ellwanger, C.~Hugonie and A.~M.~Teixeira,
  %``The Next-to-Minimal Supersymmetric Standard Model,''
  Phys.\ Rept.\  {\bf 496}, 1 (2010)
  [arXiv:0910.1785 [hep-ph]].
  %%CITATION = ARXIV:0910.1785;%%
  %531 citations counted in INSPIRE as of 19 juin 2015

%\cite{Maniatis:2009re}
\bibitem{Maniatis:2009re}
  M.~Maniatis,
  %``The Next-to-Minimal Supersymmetric extension of the Standard Model reviewed,''
  Int.\ J.\ Mod.\ Phys.\ A {\bf 25}, 3505 (2010)
  [arXiv:0906.0777 [hep-ph]].
  %%CITATION = ARXIV:0906.0777;%%
  %203 citations counted in INSPIRE as of 19 Jun 2015

%\cite{Cao:2014kya}
\bibitem{Cao:2014kya}
  J.~Cao, D.~Li, L.~Shang, P.~Wu and Y.~Zhang,
  %``Exploring the Higgs Sector of a Most Natural NMSSM and its Prediction on Higgs Pair Production at the LHC,''
  JHEP {\bf 1412}, 026 (2014)
  [arXiv:1409.8431 [hep-ph]].
  %%CITATION = ARXIV:1409.8431;%%
  %10 citations counted in INSPIRE as of 20 juin 2015

%\cite{Kang:2012sy}
\bibitem{Kang:2012sy}
  Z.~Kang, J.~Li and T.~Li,
  %``On Naturalness of the MSSM and NMSSM,''
  JHEP {\bf 1211}, 024 (2012)
  [arXiv:1201.5305 [hep-ph]].
  %%CITATION = ARXIV:1201.5305;%%
  %111 citations counted in INSPIRE as of 26 juin 2015


%\cite{Barbieri:2006bg}
\bibitem{Barbieri:2006bg}
  R.~Barbieri, L.~J.~Hall, Y.~Nomura and V.~S.~Rychkov,
  %``Supersymmetry without a Light Higgs Boson,''
  Phys.\ Rev.\ D {\bf 75}, 035007 (2007)
  [hep-ph/0607332].
  %%CITATION = HEP-PH/0607332;%%
  %137 citations counted in INSPIRE as of 27 juin 2015

%\cite{Cao:2008un}
\bibitem{Cao:2008un}
  J.~Cao and J.~M.~Yang,
  %``Current experimental constraints on NMSSM with large lambda,''
  Phys.\ Rev.\ D {\bf 78}, 115001 (2008)
  [arXiv:0810.0989 [hep-ph]].
  %%CITATION = ARXIV:0810.0989;%%
  %29 citations counted in INSPIRE as of 05 aout 2015

%\cite{Perelstein:2012qg}
\bibitem{Perelstein:2012qg}
  M.~Perelstein and B.~Shakya,
  %``XENON100 implications for naturalness in the MSSM, NMSSM, and $\lambda$-supersymmetry model,''
  Phys.\ Rev.\ D {\bf 88}, no. 7, 075003 (2013)
  [arXiv:1208.0833 [hep-ph]].
  %%CITATION = ARXIV:1208.0833;%%
  %60 citations counted in INSPIRE as of 27 juin 2015

%\cite{Hall:2011aa}
\bibitem{Hall:2011aa}
  L.~J.~Hall, D.~Pinner and J.~T.~Ruderman,
  %``A Natural SUSY Higgs Near 126 GeV,''
  JHEP {\bf 1204}, 131 (2012)
  [arXiv:1112.2703 [hep-ph]].
  %%CITATION = ARXIV:1112.2703;%%
  %373 citations counted in INSPIRE as of 27 juin 2015


%\cite{Agashe:2012zq}
\bibitem{Agashe:2012zq}
  K.~Agashe, Y.~Cui and R.~Franceschini,
  %``Natural Islands for a 125 GeV Higgs in the scale-invariant NMSSM,''
  JHEP {\bf 1302}, 031 (2013)
  [arXiv:1209.2115 [hep-ph]].
  %%CITATION = ARXIV:1209.2115;%%
  %65 citations counted in INSPIRE as of 27 juin 2015

  %\cite{Gherghetta:2012gb}
\bibitem{Gherghetta:2012gb}
  T.~Gherghetta, B.~von Harling, A.~D.~Medina and M.~A.~Schmidt,
  %``The Scale-Invariant NMSSM and the 126 GeV Higgs Boson,''
  JHEP {\bf 1302}, 032 (2013)
  [arXiv:1212.5243 [hep-ph]].
  %%CITATION = ARXIV:1212.5243;%%
  %72 citations counted in INSPIRE as of 27 juin 2015

%\cite{Barbieri:2013hxa}
\bibitem{Barbieri:2013hxa}
  R.~Barbieri, D.~Buttazzo, K.~Kannike, F.~Sala and A.~Tesi,
  %``Exploring the Higgs sector of a most natural NMSSM,''
  Phys.\ Rev.\ D {\bf 87}, no. 11, 115018 (2013)
  [arXiv:1304.3670 [hep-ph]].
  %%CITATION = ARXIV:1304.3670;%%
  %45 citations counted in INSPIRE as of 27 juin 2015

%\cite{Farina:2013fsa}
\bibitem{Farina:2013fsa}
  M.~Farina, M.~Perelstein and B.~Shakya,
  %``Higgs Couplings and Naturalness in $\lambda$-SUSY,''
  JHEP {\bf 1404}, 108 (2014)
  [arXiv:1310.0459 [hep-ph]].
  %%CITATION = ARXIV:1310.0459;%%
  %14 citations counted in INSPIRE as of 27 Jun 2015


\bibitem{higgsfield}
U. Ellwanger, C. Hugonie and A. M. Teixeira, Phys. Rept. 496, 1 (2010);
M. Maniatis, Int. J.Mod. Phys. A25 (2010) 3505;
S. F. King, P. L. White, Phys. Rev. D 52, 4183 (1995).

%\cite{Ellwanger:2011mu}
\bibitem{Ellwanger:2011mu}
  U.~Ellwanger, G.~Espitalier-Noel and C.~Hugonie,
  %``Naturalness and Fine Tuning in the NMSSM: Implications of Early LHC Results,''
  JHEP {\bf 1109}, 105 (2011)
  [arXiv:1107.2472 [hep-ph]].
  %%CITATION = ARXIV:1107.2472;%%
  %67 citations counted in INSPIRE as of 28 Jun 2015

%\cite{ATLAS2014}
\bibitem{ATLAS2014}
 ATLAS collaboration,
 % ''ATLAS interpretation of the combined measurements of coupling
 %properties of the Higgs boson in terms of its production cross sections,''
ATL-PHYS-PUB-2014-009 (2014).


%\cite{CMS2014}
\bibitem{CMS2014}
CMS collaboration,
%''Precise determination of the mass of the Higgs boson and studies of the
%compatibility of its couplings with the standard model, ''
CMS-PAS-HIG-14-009 (2014).

%\cite{Aaltonen:2013ioz}
\bibitem{Aaltonen:2013ioz}
  T.~Aaltonen {\it et al.} [CDF and D0 Collaborations],
  %``Higgs Boson Studies at the Tevatron,''
  Phys.\ Rev.\ D {\bf 88}, no. 5, 052014 (2013)
  [arXiv:1303.6346 [hep-ex]].
  %%CITATION = ARXIV:1303.6346;%%
  %128 citations counted in INSPIRE as of 05 Aug 2015

%\cite{Espinosa:2012ir}
\bibitem{Espinosa:2012ir}
  J.~R.~Espinosa, C.~Grojean, M.~Muhlleitner and M.~Trott,
  %``Fingerprinting Higgs Suspects at the LHC,''
  JHEP {\bf 1205}, 097 (2012)
  [arXiv:1202.3697 [hep-ph]].
  %%CITATION = ARXIV:1202.3697;%%
  %198 citations counted in INSPIRE as of 28 juin 2015

  %\cite{Giardino:2012ww}
\bibitem{Giardino:2012ww}
  P.~P.~Giardino, K.~Kannike, M.~Raidal and A.~Strumia,
  %``Reconstructing Higgs boson properties from the LHC and Tevatron data,''
  JHEP {\bf 1206}, 117 (2012)
  [arXiv:1203.4254 [hep-ph]].
  %%CITATION = ARXIV:1203.4254;%%
  %168 citations counted in INSPIRE as of 28 juin 2015


%\cite{Dolan:2012rv}
\bibitem{Dolan:2012rv}
  M.~J.~Dolan, C.~Englert and M.~Spannowsky,
  %``Higgs self-coupling measurements at the LHC,''
  JHEP {\bf 1210}, 112 (2012)
%  doi:10.1007/JHEP10(2012)112
  [arXiv:1206.5001 [hep-ph]].
  %%CITATION = doi:10.1007/JHEP10(2012)112;%%
  %153 citations counted in INSPIRE as of 29 janv. 2016


%\cite{McCullough:2013rea}
\bibitem{McCullough:2013rea}
  M.~McCullough,
  %``An Indirect Model-Dependent Probe of the Higgs Self-Coupling,''
  Phys.\ Rev.\ D {\bf 90}, no. 1, 015001 (2014)
  [Phys.\ Rev.\ D {\bf 92}, no. 3, 039903 (2015)]
%  doi:10.1103/PhysRevD.90.015001, 10.1103/PhysRevD.92.039903
  [arXiv:1312.3322 [hep-ph]].
  %%CITATION = doi:10.1103/PhysRevD.90.015001, 10.1103/PhysRevD.92.039903;%%
  %21 citations counted in INSPIRE as of 29 Jan 2016



 %\cite{Wu:2015nba}
\bibitem{Wu:2015nba}
  L.~Wu, J.~M.~Yang, C.~P.~Yuan and M.~Zhang,
  %``Higgs self-coupling in the MSSM and NMSSM after the LHC Run 1,''
  Phys.\ Lett.\ B {\bf 747}, 378 (2015)
%  doi:10.1016/j.physletb.2015.06.020
  [arXiv:1504.06932 [hep-ph]].
  %%CITATION = doi:10.1016/j.physletb.2015.06.020;%%
  %15 citations counted in INSPIRE as of 28 Jan 2016

%\cite{Cao:2015oaa}
\bibitem{Cao:2015oaa}
  Q.~H.~Cao, B.~Yan, D.~M.~Zhang and H.~Zhang,
  %``Resolving the Degeneracy in Single Higgs Production with Higgs Pair Production,''
  Phys.\ Lett.\ B {\bf 752}, 285 (2016)
%  doi:10.1016/j.physletb.2015.11.045
  [arXiv:1508.06512 [hep-ph]].
  %%CITATION = doi:10.1016/j.physletb.2015.11.045;%%
  %3 citations counted in INSPIRE as of 29 Jan 2016

\bibitem{Nhung:2013lpa}
  D.~T.~Nhung, M.~Muhlleitner, J.~Streicher and K.~Walz,
  %``Higher Order Corrections to the Trilinear Higgs Self-Couplings in the Real NMSSM,''
  JHEP {\bf 1311}, 181 (2013)
  [arXiv:1306.3926 [hep-ph]].
  %%CITATION = ARXIV:1306.3926;%%
  %29 citations counted in INSPIRE as of 06 juil. 2015

%\cite{Ellwanger:2013ova}
\bibitem{Ellwanger:2013ova}
  U.~Ellwanger,
  %``Higgs pair production in the NMSSM at the LHC,''
  JHEP {\bf 1308}, 077 (2013)
  [arXiv:1306.5541 [hep-ph]].
  %%CITATION = ARXIV:1306.5541;%%
  %47 citations counted in INSPIRE as of 06 juil. 2015

  %\cite{Han:2013sga}
\bibitem{Han:2013sga}
  C.~Han, X.~Ji, L.~Wu, P.~Wu and J.~M.~Yang,
  %``Higgs pair production with SUSY QCD correction: revisited under current experimental constraints,''
  JHEP {\bf 1404}, 003 (2014)
  [arXiv:1307.3790 [hep-ph]].
  %%CITATION = ARXIV:1307.3790;%%
  %25 citations counted in INSPIRE as of 06 Jul 2015

  %\cite{Cao:2013si}
\bibitem{Cao:2013si}
  J.~Cao, Z.~Heng, L.~Shang, P.~Wan and J.~M.~Yang,
  %``Pair Production of a 125 GeV Higgs Boson in MSSM and NMSSM at the LHC,''
  JHEP {\bf 1304}, 134 (2013)
  [arXiv:1301.6437 [hep-ph]].
  %%CITATION = ARXIV:1301.6437;%%
  %43 citations counted in INSPIRE as of 06 Jul 2015

%\cite{Cao:2013cfa}
\bibitem{Cao:2013cfa}
  J.~Cao, Y.~He, P.~Wu, M.~Zhang and J.~Zhu,
  %``Higgs Phenomenology in the Minimal Dilaton Model after Run I of the LHC,''
  JHEP {\bf 1401}, 150 (2014)
  [arXiv:1311.6661 [hep-ph]].
  %%CITATION = ARXIV:1311.6661;%%
  %15 citations counted in INSPIRE as of 06 Jul 2015

%\cite{Liu:2013woa}
\bibitem{Liu:2013woa}
  J.~Liu, X.~P.~Wang and S.~h.~Zhu,
  %``Discovering extra Higgs boson via pair production of the SM-like Higgs bosons,''
  arXiv:1310.3634 [hep-ph].
  %%CITATION = ARXIV:1310.3634;%%
  %22 citations counted in INSPIRE as of 06 juil. 2015

%\cite{No:2013wsa}
\bibitem{No:2013wsa}
  J.~M.~No and M.~Ramsey-Musolf,
  %``Probing the Higgs Portal at the LHC Through Resonant di-Higgs Production,''
  Phys.\ Rev.\ D {\bf 89}, no. 9, 095031 (2014)
  [arXiv:1310.6035 [hep-ph]].
  %%CITATION = ARXIV:1310.6035;%%
  %35 citations counted in INSPIRE as of 06 juil. 2015

%\cite{Barger:2014taa}
\bibitem{Barger:2014taa}
  V.~Barger, L.~L.~Everett, C.~B.~Jackson, A.~D.~Peterson and G.~Shaughnessy,
  %``New physics in resonant production of Higgs boson pairs,''
  Phys.\ Rev.\ Lett.\  {\bf 114}, no. 1, 011801 (2015)
  [arXiv:1408.0003 [hep-ph]].
  %%CITATION = ARXIV:1408.0003;%%
  %13 citations counted in INSPIRE as of 06 juil. 2015




\end{thebibliography}
\end{document}